# A Python-Based Approach to Sputter Deposition Simulations in Combinatorial Materials Science

Felix Thelen[1], Rico Zehl[1], Jan Lukas Bürgel[1], Diederik Depla[2], Alfred Ludwig[1,*]

[1] Materials Discovery and Interfaces, Institute for Materials, Ruhr University Bochum, 44801 Bochum, Germany

[2] Department of Solid State Sciences, Ghent University, 9000 Gent, Belgium

*corresponding author: alfred.ludwig@rub.de

**Abstract**

Magnetron sputtering is an essential technique in combinatorial materials science, enabling the efficient synthesis of thin-film materials libraries with continuous compositional gradients. For exploring multidimensional search spaces, minimizing preliminary experiments is essential, as numerous materials libraries are required to adequately cover the space, making it crucial to fabricate only those libraries that are absolutely necessary. This can be achieved by Monte Carlo particle simulations to model the deposition profile, e.g. by SIMTRA, which is an established package mainly designed for single cathode simulations. A strong enhancement of its capabilities is the development of a Python-based wrapper, designed to simulate multi-cathode sputter processes through parallel Monte Carlo simulations. By modeling a sputter chamber and determining the relationship between deposition power and rate for an exemplary quaternary system Ni-Pd-Pt-Ru, we achieve a match between simulated and measured compositions, with a mean Euclidean distance of 3.5%. The object-oriented design of the package allows easy customization and enables the definition of complex sputter systems. Due to parallelization, simulating multiple cathodes results in no additional simulation time. These additions extend the capabilities of SIMTRA making it applicable in combinatorial materials research.

**Introduction**

In combinatorial materials science, magnetron sputtering plays a key role for the exploration of future high-performance materials. Its main advantage is the possibility to synthesize well-defined continuous compositional gradients in form of thin-film materials libraries, but sputtering is also widely used due to relatively high deposition rates and applicability to a broad spectrum of materials. Additionally, its scalability allows an easy transfer from laboratory to industrial conditions. Sputter deposition was evaluated as being the most performant combinatorial synthesis technique [1].

Knowing the relationship between deposition conditions and the resulting thin film deposition profile as well as the composition distribution across a materials library is important to reduce the number of preliminary experiments and, therefore, minimize time and materials consumption. Traditionally, experimentalists rely on the determination of the sputter rate at specified process conditions before the actual experiment to estimate the deposition results. This can be done either by measuring the deposited thickness profile, i.e. through profilometry, or with in-situ methods like quartz crystal monitors [2]. By assuming a linear relation between deposition power and sputter rate, the deposited film thickness and the composition in one measurement area of a materials library can be estimated by taking into account the nominal thickness contributions from each co-sputtered element as well as its atomic mass and nominal bulk density [3]. However, for planning multiple experiments in one multinary composition space or when a composition should be achieved with high accuracy, a single sputter rate measurement is not sufficient.

For sputter depositions, several analytical [4,5] and numerical simulations [6–10] were reported in literature, most of them based on the Monte Carlo approach, which involves following a large number of particles through the gas phase in the vacuum chamber. While analytical models offer rapid calculation, they only allow for simple geometrical configurations. In contrast, Monte Carlo models sacrifice simulation speed for much higher model complexities and the ability to not only estimate the deposition profile, but also energy and angular distributions of arriving particles [11]. The only freely available numerical simulation which also can be used for combinatorial magnetron sputtering is SIMTRA (Simulation of Metal Transport) [12]. SIMTRA allows for a flexible geometrical configuration of the sputter chamber through user-definable 3D geometries in a graphical user interface and is available as a compiled Windows application.

Since SIMTRA is designed for the simulation of a single magnetron sputter cathode at a time, the command line version of the code was wrapped in a Python environment, enabling the definition of sputter chambers through code and executing the time-consuming Monte Carlo calculations through external software. This approach also enables parallel simulation of multiple magnetrons with multi-threading, decreasing calculation times significantly when simulating co-sputter systems with two or more magnetrons sputtering simultaneously. The functionalities and capabilities of this wrapper, titled pySIMTRA, are presented and demonstrated by comparing the simulated and measured compositions of thin films in seven multinary materials libraries in the system Ni-Pd-Pt-Ru.

**Methods**

*The SIMTRA model*

Before SIMTRA is tracking individual particles in their movement through the vacuum chamber, the initial position, launch direction and energy of the particles are sampled from multiple distributions in Monte Carlo manner. The initial position is sampled from the measured racetrack, which is specified by profilometry data in 2D or 3D [13]. For the nascent energy distribution $\phi(E)$, the Thompson distribution [14] with a cut-off energy $E_{max}$ is used:

$$\phi(E)\,dE \approx \frac{E}{(E+U_s)^3}\,dE, \qquad E_{max} = k \cdot \Lambda \cdot E_i - U_s \text{ with } \Lambda = \frac{4 \cdot M_i \cdot M_t}{(M_i + M_t)^2}$$

with $U_s$ being the surface binding energy, $\Lambda$ the maximum energy fraction loss dependent on the mass of the incident ion $M_i$ and target atom $M_t$ and $k = 0.4$ being a factor accounting for the head-on-head collisions. The energy of the incident ion $E_i$ depends on the discharge voltage [11]. For the nascent angular distribution, a linearly combined cosine distribution with six coefficients $c_i$ is defined with $\theta$ being the polar emission angle [12]:

$$\frac{d^2Y}{d^2\Omega}(\theta) = \sum_{i=0}^{5} c_i \cdot \cos^i\theta$$

For lower ion energies, the angular distribution is approximated with a cosine function ($c_1 = 1, c_{i \neq 1} = 0$) [7]. Modelling the energy and angular distributions separately is an approximation as the energy of the sputtered particles is angular-dependent; therefore, SIMTRA allows the definition of angular-dependent energy distributions through the import of SRIM simulation results (Stopping and Range of Ions in Matter) [13].

After randomly choosing the initial properties of the sputtered particles from the aforementioned distributions, the free path length $\lambda$ is calculated based on the mean free path length $\lambda_m$ with

$$\lambda = -\lambda_m \cdot \ln RN_1$$

where $RN_1$ is a uniformly distributed random number in the range $]0, 1]$. The expression for $\lambda_m$ is chosen depending on the velocity of the sputtered and gas particles, in order to account for stationary gas atoms and thermalization. For calculating the gas phase transport, it is assumed that the sputtered particles are neutral atoms, that the particles only undergo elastic collisions with neutral gas atoms and that the background gas is homogenous and in thermal equilibrium [12]. After sampling the free path length, it is determined whether the particle trajectory intersects with a surface, which leads to the deposition of the particle [11]. As many collecting surfaces as desired can be predefined in SIMTRA. The surfaces are entitled "dummy" objects to distinct them from the cathode surface. In case there is no intersection with a dummy object, the collision with argon atoms is described by calculating the scattering angle and the subsequent particle velocity. The collision with other sputtered particles is neglected due to the small concentration of sputtered particles in the background gas [13].

*The SIMTRA Python wrapper*

The structure of the Python wrapper is shown in Figure 1. It operates based on (i) a file reader capable of reading and writing the SIMTRA input text files (*input.sin*) and on (ii) a simulation object which calls the command line version of the code. A sputter system in SIMTRA formally consists of a vacuum chamber, a magnetron object (cathode) and one or multiple dummy objects. The vacuum chamber defines the boundary conditions of the Monte Carlo simulation, e.g. the chamber geometry, sputter gas, gas temperature and pressure. The magnetron object holds all other non-geometrical additional model parameters like the deposited element, the racetrack shape and a 3D representation of the magnetron object. All other components of the simulation, e.g. substrates, masks or shutters, are defined with dummy objects.

All these object types were implemented as Python objects, which can either be created from SIMTRA input files or by specifying the required object parameters from scratch. The Python wrapper extends the SIMTRA functionality by allowing the definition of multiple cathodes in a single sputter system. This is achieved by splitting up the multi-cathode system into multiple single-cathode systems. These single-cathode systems are then exported to SIMTRA input files before the simulation is executed. After completion of the simulation, the output files are automatically read and merged. Hence, an intrinsic assumption of this approach is that cross deposition between the cathode is not occurring. In our experiments, this effect is expected to be minimal due to the specific design of the magnetron sources, i.e. the cathode chimney. Calling the command line version of SIMTRA from a Python environment also enables the parallel execution of multiple simulations, therefore no additional time for simulating co-sputter processes is needed. In addition, the object-oriented structure of the wrapper allows the extension of the functionality through subclassing, therefore custom sputter systems can be defined based on existing objects.

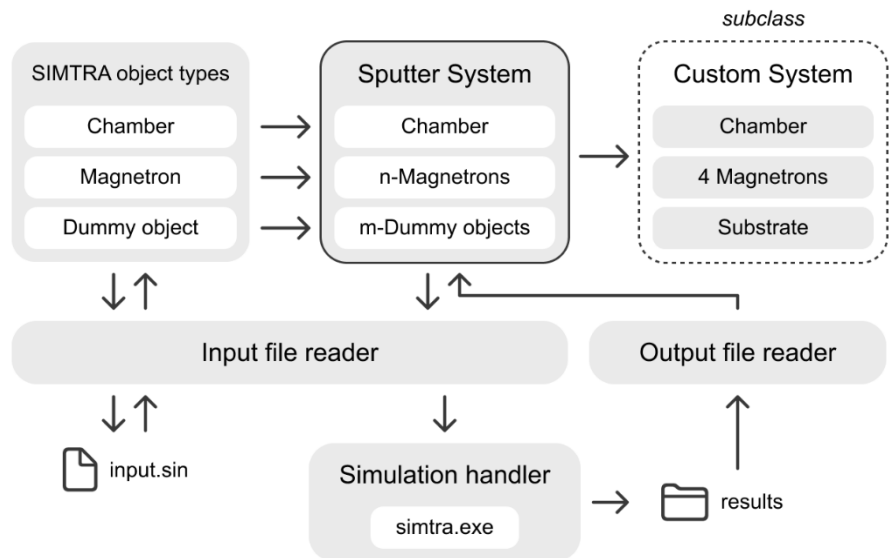

**Figure 1:** Structure of the Python wrapper. Two file readers allow the SIMTRA input files and the simulation results to be read and to convert them to Python instances. The objects can be combined to a sputter system with n-magnetrons opposed to only one in SIMTRA. Before a simulation job is started, the objects are split up into n-single magnetron systems and then passed to the SIMTRA command line version. The simulation handler is able to run multiple instances of SIMTRA in parallel. Via subclassing, the base classes can be extended to further fit to the user's needs, e.g. a standard substrate with a fixed measurement grid which was used in every experiment reported in this paper.

*Racetrack measurement*

In order to estimate the initial position of a simulated particle, SIMTRA uses the racetrack of a sputter target to define a probability function to sample from, instead of relying on the properties of the magnetic field in front of the target. To measure the racetrack of the targets in an automated way, a four-point probe resistance measurement device described in [15] was modified. It consists of a x-y-z translation stage originally designed for moving a four-point probe over a 4-inch wafer. The probe was exchanged for a single spring-loaded pin, a copper plate was placed on the substrate holder, and both were wired to a source meter. By lowering the pin onto a target positioned on the copper plate, a resistance can be measured once physical contact between pin and target is made. By logging the z-position of the pin, the racetrack profile can be inferred. Although the setup is generally capable of measuring the racetrack in 3D, a 2D profile along the diameter of the target is preferred since visual inspection of the racetrack for circular targets showed to be axially symmetric

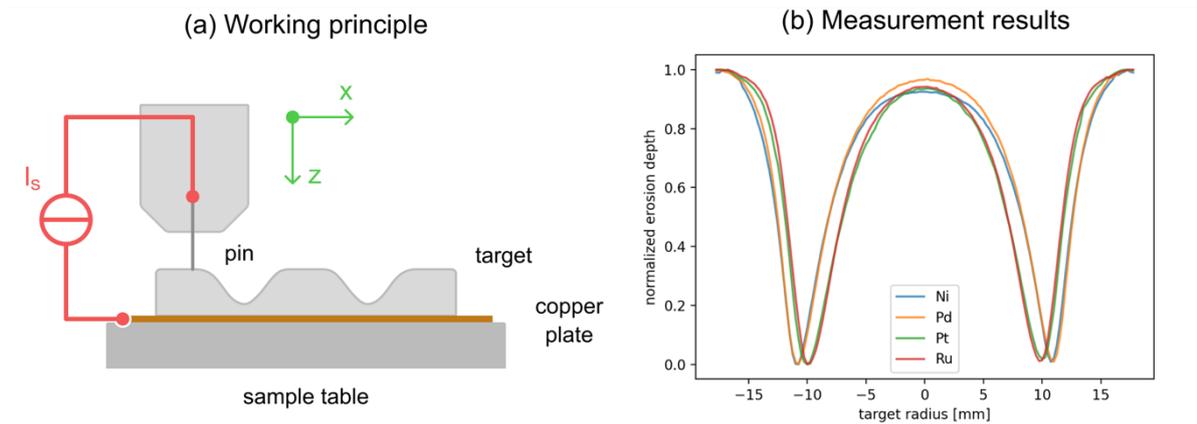

**Figure 2:** (a) Measurement approach to determine racetracks of sputter targets, (b) normalized racetrack profiles of the four used elements. In total, 16 elemental targets (1.5 inch diameter) were measured and are available with the python wrapper.

*Setup of the sputter system*

The depositions for testing the sputter simulation model were done in a co-sputter system with four cathodes (AJA International Polaris), each holding one of the following: Ni, Pd, Pt, Ru. Thin-film materials libraries were deposited on 10 cm diameter single-side polished sapphire wafers (SITUS Technicals, c-plane orientation). Three DC power supplies (2x DCXS-750, 1x DCXS-1500) and a 0313 GTC RF unit (AJA International) were used in power-control mode. The gas flow rate was set to 80 sccm Ar at a pressure of 0.5 Pa. The geometries of the chamber, the cathodes, the substrate table and the substrate were modelled in SIMTRA based on technical drawings obtained by the manufacturer. Due to the manufacturer's confidentiality rules, the actual dimensions of the model cannot be disclosed here. Figure 3 shows a photo and the modelled parts of the sputter system.

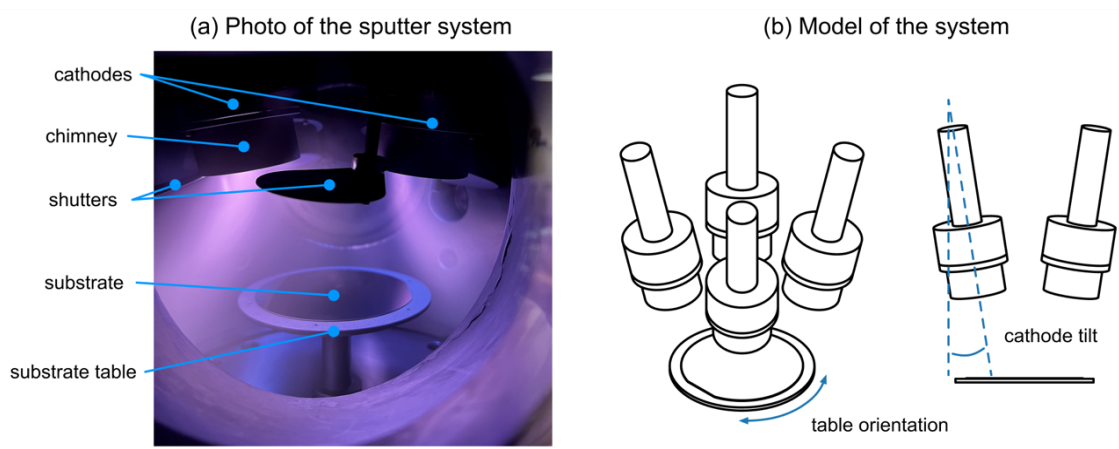

**Figure 3:** (a) Photo of the sputter system with loaded substrate during a co-deposition process using all four cathodes. The geometrical model is shown in (b). The cathode tilt can only be adjusted manually via a screw connected to a linear drive, which sets the tilt angle for all four cathodes simultaneously, without the option for individual adjustment. Also, the table orientation can be adjusted manually. Since initial tests showed no significant dependence of the cathode shutters on the simulation results, the shutters were not modelled.

SIMTRA tracks a given number of atoms through the vacuum chamber. To link the number of simulated atoms with the real number of atoms leaving the target per second, the sputter rate for each studied element needs to be known. Therefore, the sputter rates of the four elements at the respective powers were determined by deposition on crosses-patterned thermally oxidized Si wafers and subsequent film thickness profilometry measurements. The crosses were fabricated by a photolithographic process to achieve a patterned surface [16]. Before depositing the target material, a 15 nm Ta adhesion layer was fabricated in a different sputter chamber to improve adhesion of the noble metal films. After deposition on the crosses-patterned substrate, the coated wafer was washed with acetone and subsequently isopropanol to remove the crosses, resulting in well-defined regions on the substrate without a thin film. The height difference of areas with and without film was then measured with profilometry (AMBIOS XP-2 profiler), which offers a measurement repeatability of about 1 nm or 0.1% of the nominal step height, whichever is greater. For the DC generators, the samples were fabricated by depositing each element at 5 W and 35 W, for RF at 7 W and 30 or 60 W until approximately 100 nm of film accumulated on the substrate. A detailed list of the sputter parameters can be found in the supporting information. To acquire the rate at such low powers, long deposition times of up to 8 hours were necessary.

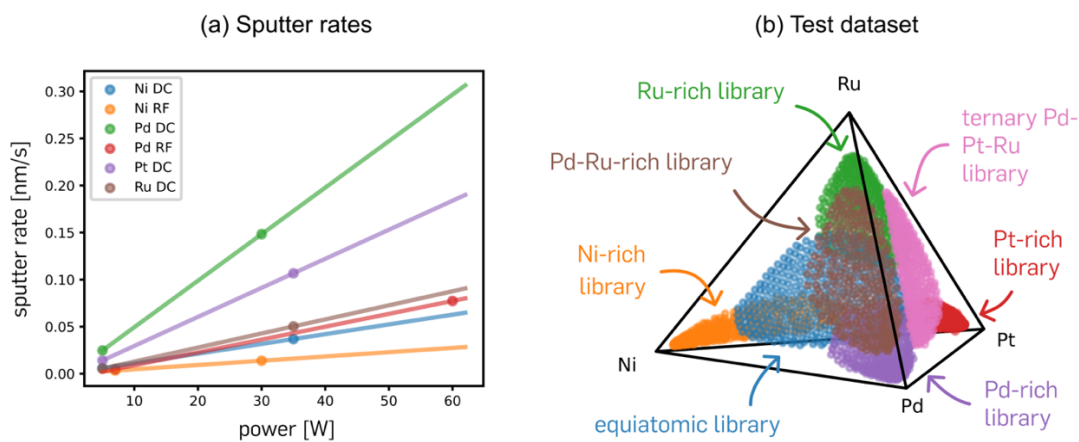

**Figure 4:** (a) Sputter rates of the four elements and their power supply pairings. The points indicate the experimentally acquired values, the lines show the assumed linear relationship of power and sputter rate. (b) Test dataset containing the compositional mappings of seven materials libraries in the Ni-Pd-Pt-Ru composition space.

The model was tested based on the compositions of seven materials libraries in the Ni-Pd-Pt-Ru system, originally fabricated for exploring quaternary composition spaces for single phase solid solutions and electrocatalytic applications [17,18]. The test dataset consisting of six quaternary and one ternary library is shown in Figure 4 (b). Due to its high sputter yield, the cathode equipped with Pd was assigned to an RF power supply after the first deposition. Table 1 shows an overview of the power parameters. The libraries were characterized in terms of their chemical composition by automatically measuring 342 areas on each library using energy-dispersive X-ray spectroscopy (EDS) in a scanning electron microscope (JEOL L 7200F) equipped with an EDS detector (Oxford AZtecEnergy X-MaxN 80 mm$^2$) with a nominal measurement accuracy of ±1 at.%. The sputter rates were determined immediately following the fabrication of the seven test libraries, as sputter rates change over the lifetime of a target.

**Table 1:** Power setpoint values used for each deposition and cathode. A full list of sputter parameters can be found in the supporting information. Except for the deposition of the equiatomic library, each cathode was paired with the same power supply for all depositions.

|  | Ni | Pd | Pt | Ru |
|---|---|---|---|---|
|  | Power [W] | | | |
| Library | DC | RF | DC | DC |
| Equiatomic | 60 (RF) | 5 (DC) | 9 | 16 |
| Ni-rich | 35 | 7 | 5 | 5 |
| Ru-rich | 5 | 7 | 5 | 35 |
| Pt-rich | 5 | 7 | 35 | 5 |
| Pd-rich | 10 | 60 | 5 | 5 |
| Pd-Ru-rich | 12 | 40 | 5 | 32 |
| Pd-Pt-Ru | - | 40 | 18 | 35 |

The calculations were performed with pySIMTRA. For the angular distribution, a cosine function was chosen. The energy distribution was modelled with the Thompson distribution using an energy cut-off equal to the discharge voltage of the DC-sources and to the RF-bias for the RF power supply. The surface binding energy was chosen by SIMTRA based on the sputtered element. The particle transport was simulated with diffusion and gas motion taken into account. The interactions of the sputtered particles with the background gas were described using a screened Coulomb potential Moliére screening function [13,19]. About $10^8$ particles were simulated per cathode. Since SIMTRA only estimates the number of particles arriving on the substrate, the fraction of arriving species of each element needs to be converted into a composition by weighting the fraction with the sputter rates. Afterwards, the compositions are multiplied by the fraction of the nominal bulk density and atomic mass in order to arrive at the unit of atomic percent. The simulation of the composition distribution of a single materials library takes approximately 20 minutes on an AMD Ryzen 7 5800X 3.8 GHz 8-Core PC.

**Results and discussion**

The simulation accuracy can be assessed based on the Euclidean distance $d_e$ of measured $c_m$ and simulated sets of compositions $c_s$ in an $n$-dimensional composition space:

$$d_e[at.\%] = \sqrt{\sum_{i=1}^{n=4}(c_{m,i} - c_{s,i})^2}$$

The maximum Euclidean distance in a quaternary composition space ($n = 4$) can be found between the pure elements and is equal to $d_e = \sqrt{2} \cdot 100 \, [at.\%]$. Therefore, the Euclidean distance can be weighted with $\sqrt{2}$ in order to arrive at a value in per-cent and is used here as an accuracy indicator.

$$d_e[\%] = \frac{d_e[at.\%]}{\sqrt{2}}$$

The Euclidean distance distributions of each materials library are shown in Figure 5 in form of violin plots. In each violin, the center dashed line denotes the mean, the lower and upper lines the quartiles of a distribution. For reference, the EDS measurement accuracy is shown by the dashed line. Across six materials libraries, the simulation achieves a mean Euclidean distance of less than 5% with a maximum deviation of 9%. Especially the equiatomic materials library close to the center of the composition space shows a high simulation accuracy.

Both the measured and simulated compositions of this equiatomic library are visualized in Figure 6 (a) emphasizing the high accuracy of the simulation. Animated versions of the composition space plots of all materials libraries can be found in the supplementary information. The similarity of the point cloud shape of the measured and simulated compositions hints at a sufficient agreement of the modelled energy and angular distribution with the real-world sputter conditions. Additionally, the overlap of the point clouds indicates a high accuracy of the sputter rates.

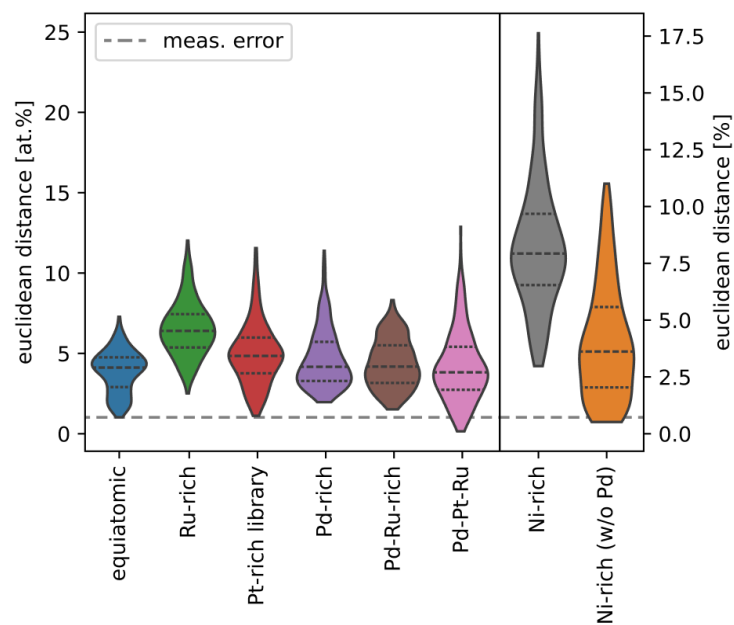

**Figure 5:** Euclidean distances (in atomic percent and percent) of the measured and simulated compositions from the seven materials libraries. The Ni-rich library was simulated twice, with and without Pd.

The only exception to the simulation performance is the Ni-rich library, which shows a mean Euclidean distance of 7.5% and a maximum deviation of 18%. When comparing both composition distributions of this library on the composition space shown in Figure 6 (b), the measured compositions suggest a considerably smaller content of Pd. While the simulated Pd content ranges from 2.2 to 3.8 at.%, the measured content only ranges around 0.9 to 1.2 at.%. This indicates a significantly smaller sputter rate of Pd during the experiment in comparison to the determined rates. When simulating the materials library without the Pd-cathode, the simulation error is reduced to a mean Euclidean distance of 3%. Figure 6 (b) also shows the simulated composition distribution without Pd which aligns much more to the measured distribution. This suggests that the origin of the deviation can be attributed to the deposition of Pd. Pd is deposited using an RF power supply at an extreme low power (7 W) which runs the discharge close to its extinction limit. It was observed that during the deposition of the library the Pd-cathode was indeed unstable during the sputter process and occasionally switched off during the deposition.

A limitation of using the Euclidean distance as an accuracy measure is its sensitivity to alignment between the measurement and simulation grids. Since the substrate table can have a rotational offset inside the vacuum chamber, which can only be aligned by visual inspection before closing the chamber, there is a risk of misalignment between the simulation and measurement grids. This rotational offset can cause the Euclidean distance distributions shown in Figure 5 to be stretched, which may lead to the accuracy of the simulation appearing worse than it actually is. Therefore, while the Euclidean distance provides an effective, condensed metric for an overall accuracy assessment, it may overstate the discrepancies due to this rotational misalignment. In order to assess the simulation accuracy without the dependence of the measurement grid, the elemental content distributions of each simulated and measured materials library are compared in Figure 7. This approach preserves an additional dimension of the data and therefore allows the analysis of the element-wise deviation of the simulation as well.

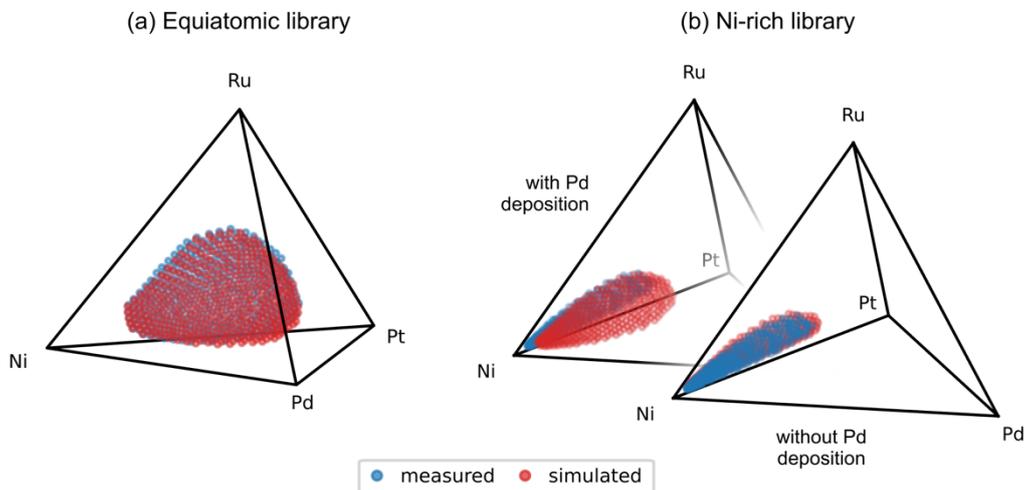

**Figure 6:** Comparison of (a) the equiatomic and (b) Ni-rich library on the composition space. The composition of the Ni-rich library is shown with and without the deposition of the Pd cathode. The deviation of measured and simulated composition is decreased when the Pd deposition is not taken into account, hinting at an unstable deposition.

Ideally, the simulated elemental distributions should be mirrored replicas of the measured distribution. Similar to Figure 5, the library with the highest simulation accuracy is the equiatomic library, see Figure 7 (a), as the composition distributions of all elements match closely between measurement and simulation. For the libraries with a high content of one element, Figure 7 (b-e), the simulated composition distributions appear shifted compared to the measured ones. This can be explained by inaccurately measured sputter rates as the ratio of sputter rates affects the position of the point cloud in the composition space. Some rates were determined from films with a thickness smaller than 100 nm due to the low deposition rate at 5-7 W and the high monetary costs of consuming noble metals. With a thickness measurement error of 1 nm, this results in an uncertainty of the sputter rates of higher than 1%.

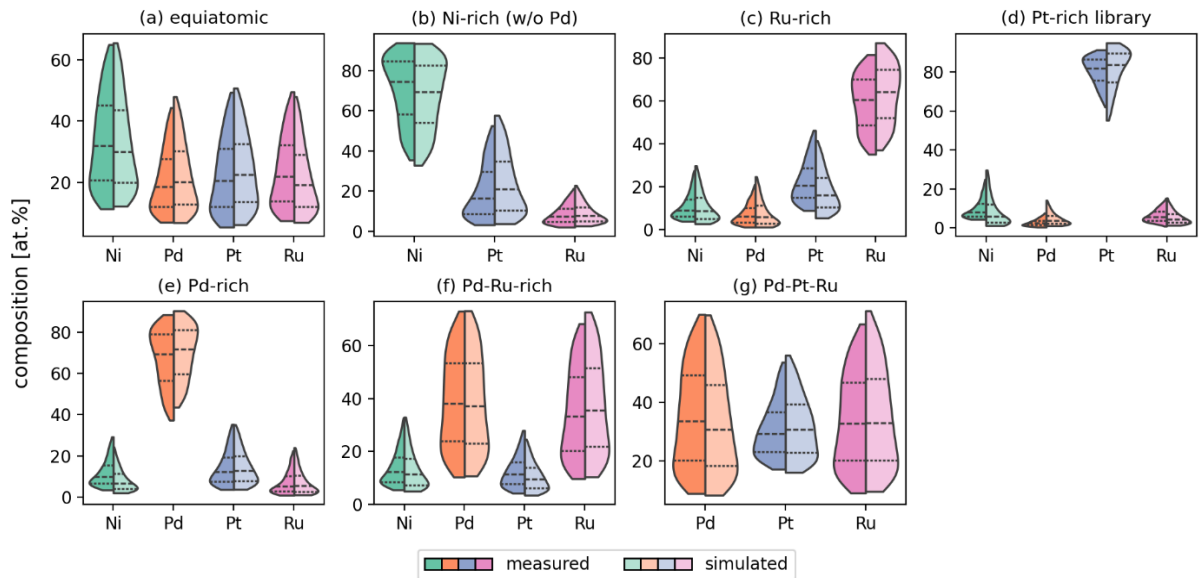

**Figure 7:** Comparison of the element-wise measured (dark colors) and simulated (light colors) composition distributions.

In contrast to the quaternary libraries, the simulated composition distributions of the ternary library, shown Figure 7 (g), exhibit a wider range compared to the measured compositions. This discrepancy could be explained by an incorrectly assumed cathode tilt. The deposition angle of the cathodes of the used sputter system can be adjusted manually by a linear drive connected to a tilting mechanism as shown in Figure 3 (b). However, this tilt adjustment applies to all cathodes simultaneously rather than individually. Slight changes to this tilt can already occur unintentionally while opening and closing the chamber for a sample exchange and can have a rather large effect due to the non-linearity of the relationship between cathode tilt and scale on the adjustment screw. An increased tilt results in a smaller compositional range of the materials library, explaining the simulation offset of the ternary library.

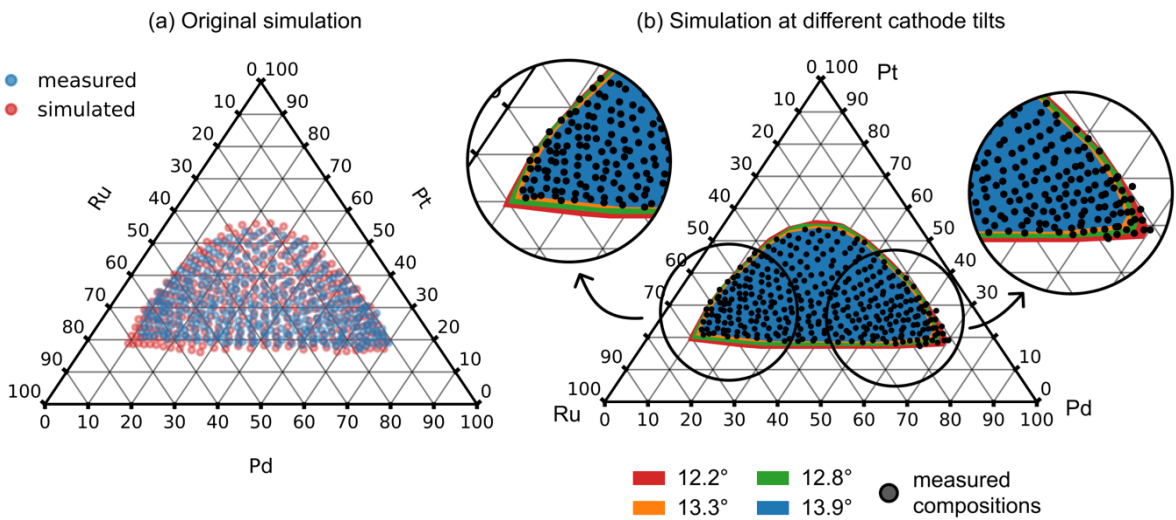

**Figure 8:** Measured and simulated compositions of the ternary Pd-Pt-Ru library on the composition space (a). The measured distribution shows a smaller range of compositions, hinting at a wrongly assumed cathode tilt. The deviation was further investigated with (b), which show the covered composition space of the simulations done at various cathode tilts.

This influence is confirmed by Figure 8 (a), which shows the compositional range covered by both the simulated and measured materials libraries. Here, the measured compositions span a 4.3% smaller compositional area compared to simulation. In order to estimate the cathode tilt deviation, the simulation was repeated with varying tilts based on the scale on the adjustment screw. The results, visualized in Figure 8 (b), depict the compositional area covered by each simulated materials library. By increasing the tilt from the original 12.2° to 12.8°—equivalent to a 0.5 mm decrease on the adjustment screw—the simulated compositional area aligns more closely with the measured area. Furthermore, as highlighted in the zoomed-in section on the right, the deviation is not only influenced by the cathode tilt, but also by a deviation in sputter rates, as the measurements suggest a higher content of Pd compared to the simulation. When increasing the tilt further to 13.3° or 13.9° respectively, the Pd content is underestimated too much by the simulation, suggesting that the sputter process was likely run with a cathode tilt deviation of 0.5°. Overall, maintaining accurate cathode tilt is crucial for ensuring simulation accuracy. However, for practical purposes, a tilt accuracy of approximately 1 degree should be sufficient to achieve reliable simulation results.

In addition to inaccuracies of the measured sputter rates and the cathode tilt influencing the simulation accuracy, the thin-film density can have a significant effect as well, both on the shift and range of the compositional distributions. As the density of magnetron sputtered thin films generally is dependent on many parameters such as the process pressure, the target-to-substrate-distance, temperature, as well as the deposition power, the film density is likely different for different sputter conditions and even across the surface of a materials library. Therefore, different and unknown thin-film density values during the sputter rate determination and fabrication of the materials libraries can further decrease the simulation accuracy, since the conversion from the ratio of arriving particles to the composition relies on the nominal bulk density of the sputtered elements. For future studies, this influence could be avoided by using methods of rate determination not relying on film density.

Since the simulation assumes the independence of the energy and angular distributions, which is only an approximation of real-world sputter mechanisms, using additional codes like SDTrimSP [20] could result in more accurate composition predictions. Due to the higher simulation time and computational effort, this factor was not considered in this study.

**Conclusions**

The developed Python wrapper for SIMTRA enables fast and efficient simulation of co-sputtering processes by allowing the parallel execution of the magnetron sputter Monte Carlo simulation. Simulations can be completed in 20-40 minutes depending on geometry complexity, and multiple cathodes (2-8) can be simulated in parallel depending on hardware specifications. Therefore, the simulation requires significantly less time than conducting the equivalent experimental deposition. The object-oriented structure of the wrapper also supports straightforward extensions with custom classes.

Testing the wrapper on seven materials libraries in the Ni-Pd-Pt-Ru composition space demonstrated that it provides reliable predictions of the compositional distributions. By estimating the relationship between deposition power and sputter rates through thickness profilometry and modeling the chamber geometry accurately in SIMTRA, a mean Euclidean distance of 3.5% between simulated and measured compositions was achieved. This accuracy makes the simulation well-suited for determining process parameters in advance, thereby reducing the need for preliminary experiments and saving time in the setup process.

The comparison of simulated and experimental data also revealed valuable insights: deviations between simulated and measured compositions can indicate geometric issues, such as rotational offsets or incorrect cathode tilt, highlighting the potential for simulation to diagnose process errors. However, accurate predictions require precise calibration of sputter rates. Paired with improved methods for density-independent sputter rate estimation, this Monte Carlo simulation approach is a promising step towards creating digital twins for combinatorial materials libraries and enhancing efficiency in materials development.

**Data availability**

The dataset of this publication is accessible at Zenodo under [https://doi.org/10.5281/zenodo.14185579](https://doi.org/10.5281/zenodo.14185579). PySIMTRA was published on GitHub under [https://github.com/felixthln/pysimtra](https://github.com/felixthln/pysimtra).

**Author contributions**

F. Thelen: conceptualization, measurement, data curation, formal analysis, investigation, software, validation, visualization, supervision, writing – original draft. R. Zehl: conceptualization, supervision, writing – review & editing. J. L. Bürgel: sample fabrication, measurement, writing – review & editing. D. Depla: supervision, writing – review & editing. A. Ludwig: resources, writing – review and editing, supervision, project administration.

**Conflict of interest**

There are no conflicts to declare.

**Acknowledgements**

This work was partially supported from different projects. A. Ludwig, F. Thelen and J. L. Bürgel acknowledge funding from the European Union through the ERC Syn. G. project 101118768 (DEMI). Funding by the Deutsche Forschungsgemeinschaft (DFG, German Research Foundation) CRC TR 103, project number 190389738 and CRC 1625, project number 506711657, subproject A02 is acknowledged by R. Zehl and A. Ludwig.

## Supporting Information

**Table S1:** Sputter parameters and determined sputter rates. All rate samples were sputtered at a process pressure of 0.5 Pa and 80 sccm Ar flow on a pre-deposited 15 nm Ta adhesion layer. The thickness was determined by averaging ten manually conducted profilometry measurements. The profiler (AMBIOS XP-2) features a nominal measurement repeatability of 1 nm.

| Element | Power type | Power in W | Voltage in V | Deposition time in s | Average thickness in nm | Rate in nm/s |
|---|---|---|---|---|---|---|
| Ni | DC | 5  | 289 | 28 800 | 165.5 | 0.005745 |
|    | DC | 35 | 294 | 27 60  | 101.7 | 0.03685  |
|    | RF | 7  | 25  | 28 800 | 95.7  | 0.003323 |
|    | RF | 30 | 33  | 7 200  | 99.3  | 0.0138   |
| Pd | DC | 5  | 294 | 5 400  | 133.4 | 0.0247   |
|    | DC | 30 | 299 | 1 200  | 178.0 | 0.1483   |
|    | RF | 7  | 29  | 14 400 | 70.0  | 0.004861 |
|    | RF | 60 | 49  | 1 080  | 83.5  | 0.07731  |
| Pt | DC | 5  | 306 | 9 960  | 140.5 | 0.01411  |
|    | DC | 35 | 323 | 960    | 102.4 | 0.1067   |
| Ru | DC | 5  | 262 | 18 000 | 97.6  | 0.005422 |
|    | DC | 35 | 272 | 1 800  | 90.5  | 0.05028  |

**Table S2:** Sputter parameters and determined sputter rates. All rate samples were sputtered at a process pressure of 0.5 Pa and 80 sccm Ar flow.

| Library | Cathode | Element | Power type | Power in W | Voltage in V | Deposition time in s |
|---|---|---|---|---|---|---|
| equiatomic | 1 | Pt | DC | 9 | 305 | 7 200 |
| | 2 | Pd | DC | 5 | 270 | |
| | 3 | Ni | RF | 60 | 54 | |
| | 4 | Ru | DC | 16 | 262 | |
| Ni-rich | 1 | Pt | DC | 5 | 311 | 1 200 |
| | 2 | Pd | RF | 7 | 32 | |
| | 3 | Ni | DC | 35 | 312 | |
| | 4 | Ru | DC | 5 | 282 | |
| Ru-rich | 1 | Pt | DC | 5 | 305 | 1 200 |
| | 2 | Pd | RF | 7 | 33 | |
| | 3 | Ni | DC | 5 | 300 | |
| | 4 | Ru | DC | 35 | 281 | |
| Pt-rich | 1 | Pt | DC | 35 | 330 | 900 |
| | 2 | Pd | RF | 7 | 32 | |
| | 3 | Ni | DC | 5 | 303 | |
| | 4 | Ru | DC | 5 | 289 | |
| Pd-rich | 1 | Pt | DC | 5 | 295 | 900 |
| | 2 | Pd | RF | 60 | 55 | |
| | 3 | Ni | DC | 5 | 290 | |
| | 4 | Ru | DC | 5 | 276 | |
| Pd-Ru-rich | 1 | Pt | DC | 5 | 298 | 600 |
| | 2 | Pd | RF | 40 | 49 | |
| | 3 | Ni | DC | 12 | 233 | |
| | 4 | Ru | DC | 32 | 279 | |
| Pd-Pt-Ru | 1 | Pt | DC | 18 | 207 | 480 |
| | 2 | Pd | RF | 40 | 49 | |
| | 4 | Ru | DC | 35 | 277 | |

**Figures S3-S11:** Animated versions of the composition space plots showing the measured (blue) and simulated compositions (red) of each materials library. These are available together with the rest of the dataset in Zenodo under https://doi.org/10.5281/zenodo.14185579.